\documentclass[rnote]{aa}
\usepackage{txfonts}
\usepackage{natbib}
\usepackage{amssymb}
\bibpunct{(}{)}{;}{a}{}{,}
\usepackage{epsf}
\usepackage{graphicx}
\usepackage{epsfig}
\usepackage{graphics}
\usepackage{subfigure}
\usepackage[english]{babel}
\usepackage{rotating}
\usepackage{color}

\newcommand{\teff}{$T_{\rm{eff}}$}
\newcommand{\logg}{$\log g$}
\newcommand{\lL}{\ifmmode \log \frac{L}{L_{\sun}} \else $\log \frac{L}{L_{\sun}}$\fi}

\newcommand{\vsini}{$V$~sin$i$}

\newcommand{\vmac}{$v_{\rm mac}$}

\newcommand{\kms}{km~s$^{-1}$}
\newcommand{\msun}{M$_{\sun}$}

\begin{document}

\title{Surface abundances of OC supergiants\thanks{Based on observations obtained at the ESO/La Silla Observatory under programs 081.D-2008, 083.D-0589, 089.D-0975.}}
\author{F. Martins\inst{1}
\and S. Foschino\inst{1}
\and J.-C. Bouret\inst{2}
\and R. Barb\'a\inst{3}
\and I. Howarth\inst{4}
}
\institute{LUPM, Universit\'e de Montpellier, CNRS, Place Eug\`ene Bataillon, F-34095 Montpellier, France  \\
\and
Aix Marseille Universit\'e, CNRS, LAM (Laboratoire d'Astrophysique de Marseille) UMR 7326, 13388, Marseille, France \\
\and
Departamento de F\'{\i}sica y Astronom\'{\i}a, Universidad de La Serena, Cisternas 1200 N, La Serena, Chile \\
\and
Department of Physics \& Astronomy, University College London, Gower St, London WC1E 6BT, UK
}

\offprints{Fabrice Martins\\ \email{fabrice.martins@umontpellier.fr}}

\date{Received / Accepted }

\abstract
{Some O and B stars show unusually strong or weak lines of carbon and/or nitrogen. These objects are classified as OBN or OBC stars. It has recently been shown that nitrogen enrichment and carbon depletion are the most likely explanations for the existence of the ON class.}
{We investigate OC stars (all being supergiants) to check that surface abundances are responsible for the observed anomalous line strengths.}
{We perform a spectroscopic analysis of three OC supergiants using atmosphere models. A fourth star was previously studied by us. Our sample thus comprises all OC stars known to date in the Galaxy. We determine the stellar parameters and He, C, N, and O surface abundances.}
{We show that all stars have effective temperatures and surface gravities fully consistent with morphologically normal O supergiants. However, OC stars show little, if any, nitrogen enrichment and carbon surface abundances consistent with the initial composition. OC supergiants are thus barely chemically evolved, unlike morphologically normal O supergiants. }
{}


\authorrunning{Martins et al.}
\titlerunning{Abundances of OC stars}

\maketitle

\section{Introduction}
\label{s_intro}

Some OB stars show lines of carbon and nitrogen with unusual morphologies. \citet{jj67} were the first to report very weak nitrogen optical lines in two supergiants. Subsequent observations of OB stars revealed a number of objects that had similar spectral properties and also showed strong carbon lines \citep[][]{walborn70,jj74}. Some OB stars displayed strong nitrogen lines and weak carbon lines, too. \citet{walborn71} introduced the spectral types OBN and OBC, by analogy with Wolf-Rayet stars where similar line characteristics are seen \citep[see also][]{walborn76}. The number of OBN/OBC stars is small: according to the summary of \citet{walborn11}, only thirteen (four) Galactic O stars have an ON (OC) spectral type.

In Wolf-Rayet stars, the strength of carbon and nitrogen lines is related to their evolutionary status. Winds have removed the external layers and exposed inner regions where H and He burning products are present. \citet{walborn76} lists the possible mechanisms explaining the appearance of OBN/OBC stars. Chemical enrichment/depletion is a plausible explanation. Recently, \citet{on15} have quantitatively studied the surface abundances of ON stars \footnote{A summary of previous quantitative studies of ON stars can be found in \citet{on15}}. They concluded that strong nitrogen enrichment and carbon depletion was the reason for their spectral morphology. The physical origin of this chemical pattern remains unclear though.

In this follow-up study, we focus on OC stars. \citet{mimesO} analyzed a large sample of O stars including one OC object (HD~152249). The latter turned out to show little nitrogen enrichment, unlike similar objects without the ``C'' qualifier. Here, we study the three additional OC stars of the sample of \citet{walborn11}.

\section{Observations and sample}
\label{s_obs}

The spectroscopic data for HD~104565, HD152424, and HD~154811 were
obtained in the context of the OWN project \citep{barba10,barba14}
conducted at ESO/La Silla with the FEROS instrument mounted on the 2.2m Max-Planck telescope
(Table\ \ref{tab_obs}). They have a resolving power of 48\,000. They
were reduced by the fully automated pipeline distributed by ESO. The
spectrum of HD~152249 was acquired with the ESPaDOnS
spectropolarimeter mounted on the Canada--France--Hawaii
Telescope. Its analysis was presented in \citet{mimesO}, and we adopt
the parameters obtained in this study.

\begin{table}
\begin{center}
\caption{Observational information.} \label{tab_obs}
\begin{tabular}{lccc}
\hline
Star        & Sp.T.$^1$    &  Instrument & Date of observation\\    
            &           & \\
\hline
HD~104565   & OC9.7 Iab & FEROS      & 12 may 2008 \\
HD~152249$^2$& OC9 Iab   & ESPaDOnS   & 14 jul 2011 \\
HD~152424   & OC9.2 Ia  & FEROS      & 05 apr 2015 \\      
HD~154811   & OC9.7 Ib  & FEROS      & average  \\ 
\hline
\end{tabular}
\tablefoot{1- Spectral types are from \citet{sota11,sota14}. 2- Star analyzed by \citet{mimesO}.}
\end{center}
\end{table}

\section{Modeling and spectroscopic analysis}
\label{s_mod}

We used the atmosphere code CMFGEN \citet{hm98} to analyze the surface properties of the OC stars. The method we used is the same as that presented in \citet{on15}, and we refer the reader to this paper for a full description. CMGFEN computed non-LTE, spherically extended atmosphere models that include line-blanketing. Synthetic spectra calculated from the atmosphere models are compared to observations to determine the stellar parameters.
The main parameters were obtained as follows:

\begin{itemize}

\item \emph{Rotation and macroturbulence:} we used the Fourier-transform of \ion{O}{iii}~$\lambda$5592 to determine the projected equatorial rotational velocity \vsini\ \citep{gray76}. We then convolved a synthetic spectrum (with \teff\ and \logg\ estimated from spectral type) with a radial--tangential profile. The comparison of this synthetic spectrum with the observed profile yielded the macroturbulent velocity \vmac.

\item \emph{Effective temperature:} the ionization-balance method was used to constrain \teff, taking advantage of the numerous \ion{He}{i} and \ion{He}{ii} lines present in the optical spectra. The ionization balance of carbon (based on \ion{C}{iii} lines and \ion{C}{ii}~4267) is consistent -- within the error bars -- with the effective temperature determined from helium lines.

\item \emph{Surface gravity:} \logg\ was determined from the width of the Balmer lines wings. Broader wings are observed for higher surface gravities.

\end{itemize}

Once these parameters were constrained, we ran models with different surface abundances (for He, C, N, and O). For each star and each element, a set of clean lines was selected from the observed spectrum. These lines were quantitatively compared to the synthetic spectra by means of a  $\chi^2$ analysis from which we derived the surface abundance and associated uncertainty \citep[see][]{mimesO,on15}. Abundance determinations were performed assuming a microturbulent velocity of 10 \kms.

\section{Results}
\label{s_res}

\begin{table*}
\begin{center}
\caption{Parameters of the sample stars.} \label{tab_param}
\begin{tabular}{lcccccccccc}
\hline
Star        & Spectral   & Teff &   logg & logg$_c$&  \vsini\  &  \vmac  &   C/H         &    N/H      & O/H   \\    
            &  type      & [kK] &        &         &  [\kms]   &  [\kms] & [10$^{-4}$]  &  [10$^{-4}$] &  [10$^{-4}$] \\
\hline
HD~104565    & OC9.7 Iab  & 28  & 3.00  & 3.01 & 60   & 90 & $>$2.0 & 2.0$^{+0.7}_{-0.7}$ & $>$1.7 \\ 
HD~152249    & OC9 Iab    & 31  & 3.25  & 3.25 & 43   & 48 & 2.8$^{+0.4}_{-0.4}$ & 1.3$^{+0.7}_{-0.7}$ & $>$7.0 \\
HD~152424    & OC9.2 Ia   & 29  & 3.10  & 3.11 & 60   & 90 & 2.0$^{+1.4}_{-0.8}$ & 0.9$^{+0.2}_{-0.4}$ & $>$2.0 \\ 
HD~154811    & OC9.7 Ib   & 28  & 3.10  & 3.12 & 110  & 65 & 2.0$^{+0.8}_{-0.5}$ & 1.2$^{+1.0}_{-0.6}$ &  4.6$^{+1.4}_{-1.1}$\\
\hline
\end{tabular}
\tablefoot{Uncertainties on \teff, \logg, \vsini, and \vmac\ are $\sim$1.5kK, 0.15 dex, 10, and 20 \kms\ respectively. logg$_c$ is the surface gravity corrected for centrifugal acceleration. Abundances are number ratios. Values for HD~152249 are from \citet{mimesO}.}
\end{center}
\end{table*}

\begin{figure}[t]
\centering
\includegraphics[width=9cm]{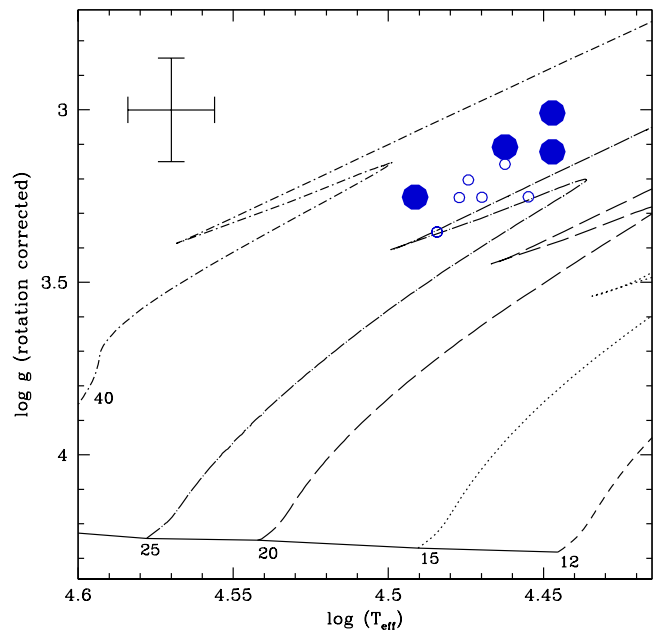}
\caption{\logg\ - log(\teff) diagram for the sample stars (filled circles) and comparison late-type supergiants (O9-9.7I - open circles). Typical uncertainties are shown in the upper-left corner. Evolutionary tracks
from \citet{ek12} including rotation are overplotted, labeled by initial masses.}
\label{fig_hr}
\end{figure}

The best fits to the optical spectra of the target stars are presented in Appendix \ref{ap_fit}. The stellar parameters and surface abundances are summarized in Table \ref{tab_param}. For all stars, we found that He/H=0.1 led to good fits (Figs.\ \ref{fit_104565} to\ \ref{fit_154811}). Figure\ \ref{fig_hr} shows the \logg\ -- \teff\ diagram with evolutionary tracks from \citet{ek12}. We added the morphologically normal late-O supergiants of \citet{mimesO} for comparison. All stars are located at, or just beyond, the terminal age main sequence. They are relatively clustered with initial masses between $\sim$20 and $\sim$40 \msun. OC and normal O supergiants are located at the same position of the diagram. As a result, OC stars do not have peculiar effective temperatures or surface gravities.

\begin{figure}[t]
\centering
\includegraphics[width=9cm]{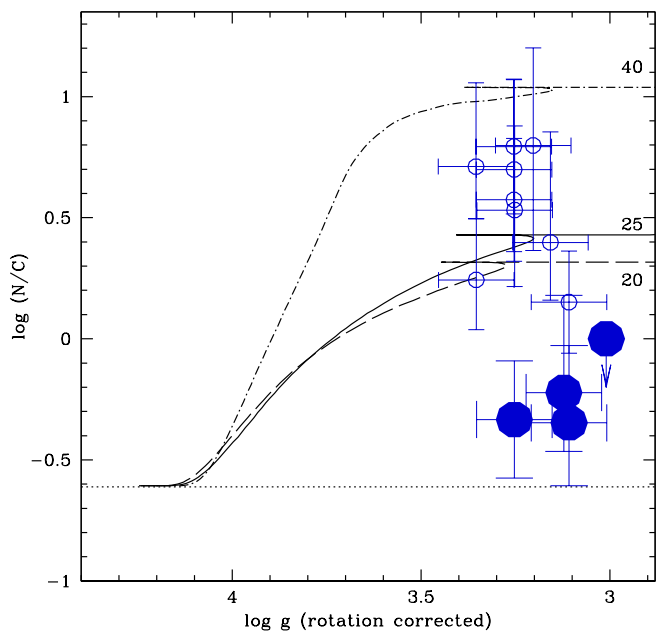}
\caption{log(N/C) - \logg\ diagram for the sample stars (filled circles) and comparison late-type supergiants (open circles). Evolutionary tracks including rotation from \citet{ek12} are overplotted, labeled by initial masses. The dotted line shows the initial chemical composition. For non-rotating models, N/C remains at the initial value (dotted line) throughout the main sequence and early post-main sequence phases.}
\label{ab_g}
\end{figure}

\begin{figure}[t]
\centering
\includegraphics[width=9cm]{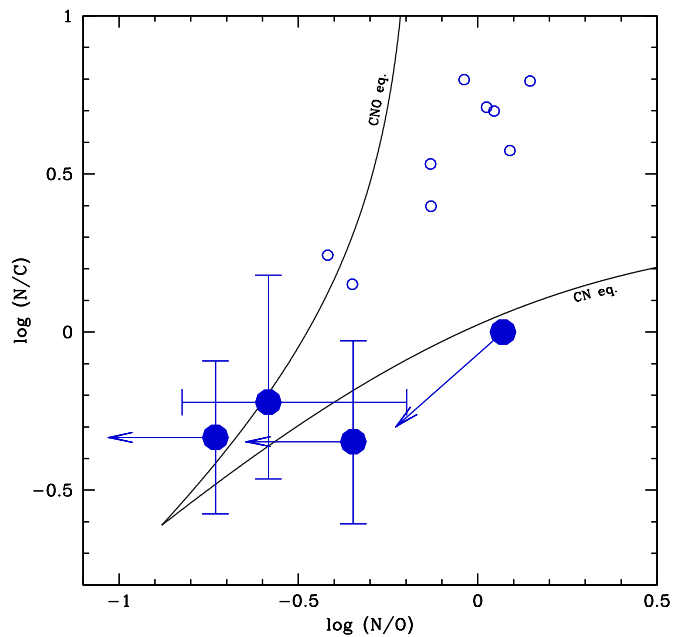}
\caption{log(N/C) - log(N/O) diagram for the sample stars (filled circles) and comparison late-type supergiants (open circles) from \citet{mimesO}. Solid lines indicate the prediction of nucleosynthesis through the partial CN or complete CNO cycle.}
\label{nco}
\end{figure}

Figure\ \ref{ab_g} shows the ratio of nitrogen-to-carbon abundance as a function of surface gravity. The evolutionary tracks of \citet{ek12} for the range of masses deduced from Fig.~\ref{fig_hr} are overplotted. The difference between normal and OC stars is striking. While normal O supergiants are accounted for relatively
well by the rotating-star evolutionary tracks, the OC stars have N/C ratios lower by 0.6-0.8 dex (factor 4 to 6). In fact, OC stars are almost consistent with no chemical processing (N/C  0.0 to 0.3 dex higher than the initial value). Figure\ \ref{nco} shows the position of the OC stars (and comparison supergiants) in the log(N/C) - log(N/O) diagram. OC stars are fully consistent with stars having experienced very little processing through the CN or CNO cycles. Morphologically normal supergiants are more chemically evolved.

\section{Discussion}

Figure\ \ref{ab_g} reveals that OC supergiants are much less chemically evolved than morphologically normal O supergiants. They are also less evolved than expected for stars of initial mass 20--40 \msun, assuming a standard initial rotational velocity of $\sim$300 \kms. Indeed, the evolutionary tracks of \citet{ek12} with this initial rotation predict values of log(N/C) between 0.6 and 1.0, while OC supergiants have log(N/C) around -0.3. It is conceivable that OC supergiants are the descendants of stars rotating very slowly on the main sequence. In that case, single-star evolution predicts that N/C should remain constant down to \logg\ $\sim$ 0.0, corresponding to the red supergiant phase. In Fig.\ \ref{ab_g}, such an evolution would be represented by the dotted line. If this explanation holds, OC supergiants would represent the descendents of main-sequence stars at the low-velocity tail of the initial rotational velocity distribution. Since the latter is continuous \citep{ramirez13}, one would expect to find O supergiants with log(N/C) over the full range $-0.3 - 1.0$ in the log(N/C) -- \logg\ diagram \citep[Fig. 9 of][]{brott11}. In Fig.\ \ref{ab_g}, there seems to be a deficit of stars for log(N/C) between -0.2 and 0.2 dex. However, the number of stars in this diagram remains small. Interestingly, \citet{hil03} studied a O7 supergiant and a OC7.5 giant in the SMC. From the stellar properties and surface abundances, they concluded that the latter was a slow rotator that did not experience strong mixing.

If slow rotation is the explanation for the chemical appearance of OC stars, their initial masses should be slightly \textit{\emph{higher}} than those of morphologically normal O supergiants. Indeed, evolutionary tracks without rotation are shifted to the right in Fig.\ \ref{fig_hr} compared to tracks including rotation. Consequently, higher masses are needed to reproduce the position of OC stars (40--60 \msun\ instead of 20--40 \msun). OC stars may thus differ from morphologically normal stars in that respect.

Slow rotation does not explain all the properties of OC stars. In particular, HD~154811 has a present-day rotational velocity of 110 \kms. This value is relatively large for O supergiant: \citet{sergio14} report \vsini\ for O9-9.7 supergiants in the range 48--132 \kms, with an average value of 88$\pm$30 \kms. Given that the projected rotational velocity drops significantly when a star leaves the main sequence, one would expect HD~154811 to have an initial \vsini\ larger than 110 \kms. For instance under the assumption of angular momentum conservation and solid-body rotation, the ratio of initial to present \vsini\ is equal to the ratio of present to initial stellar radius. Assuming a factor of three for the latter ratio \citep[according to the tracks of][]{ek12}, the initial rotational velocitiy of HD~154811 would have been 330 \kms, a fairly high value. For the other targets, the initial projected rotational velocity would have been 120--160 \kms.

The OBN/OBC classification was introduced by analogy to Wolf-Rayet stars \citep{walborn71}. In WC stars, the appearance of strong lines of carbon are caused by the large surface content of carbon produced by helium burning. One could speculate that the same is true for OC stars. However, in that case there should be no hydrogen at the surface of the stars \citep{paul07}. In addition, helium should be the dominant element, which is clearly not the case. As a result, the abundance patterns of OC stars are not due to helium-burning products.  

Surface abundances can be altered by mass transfer in binary systems. \citet{br78} investigated binarity among OC stars. Their results indicated a low binary fraction, although results were not conclusive for some stars. \citet{levato88} revisited this question and reached more robust conclusions: OC stars usually do not show the radial velocity variations typical of spectroscopic binaries. \citet{sana08} obtained time series spectroscopy of HD~152249 and did not detect any sign of a companion. The interferometric search for companions around O stars by \citet{sana14} confirms that HD~152249, HD~152424, and HD~154811 do not have bright companions. The ongoing OWN spectroscopic survey of O stars confirm these results: only HD~152424 shows radial velocity variations greater than 5 \kms. HD~152424 may be a SB1 with a low mass component (Barb\'a et al., in prep). All other OC stars are probably single objects. Consequently, the chemical abundances of OC stars are unlikely to be due to binary interaction. Besides, if mass transfer can explain nitrogen enrichment and carbon depletion, it cannot explain that surface abundances are almost unaltered. This would have to be due to tidal interaction that would suppress internal mixing processes.

Interestingly, \citet{bouret08} report very little nitrogen enrichment and almost solar carbon abundance in the star $\zeta$~Ori. \citet{sota14} classify it as ``Nwk'', meaning that the nitrogen lines are weak, as in OC stars. And $\zeta$~Ori is known to host a weak magnetic field \citep{bouret08,blazere15}, which is quite different from the strong dipoles usually observed in magnetic O stars \citep[e.g.,][]{wade15}. \citet{mimesO} showed that the chemical properties of O stars with dipole fields did not significantly differ from stars without magnetic fields. One may wonder whether for the case of a weak, small scale magnetic field, the situation is different. \citet{meynet11} show that massive stars with magnetic field of only a few Gauss may slow down very fast provided they are in solid-body rotation. In that case, there is no time for significant mixing, and stars appear chemically unevolved \citep[see Fig.~1 of][]{meynet11}. A deep spectrolarimetric study of OC stars is needed to investigate the role of magnetism.   

All OC stars in the Galaxy are late-type supergiants. Since the OC phenomenon is due to surface abundances, one can understand that it is best seen in supergiants for which chemical processing is already strong in morphologically normal stars. The surface abundances of OC supergiants are similar to those of O-normal dwarfs, probably because they are relatively chemically unevolved. Thus, the differential effect leading to the classification as OC (weak nitrogen lines/strong carbon lines) does not exist among dwarfs that are only slightly chemically evolved (\citet{mimesO}. OC stars may be seen among giants \citep{hil03}. A recent study by \citet{evans15} assigns a spectral type OC to two early type O giants in the LMC, based on the weakness of nitrogen lines. 

Regarding spectral type, there is a priori no reason for the OC phenomenon to be restricted to late-type O stars. \citet{walborn10} introduced the ``Ofc'' classification to characterize O stars with \ion{C}{iii}~4650 of the same strength as \ion{N}{iii}~4640. These features are seen in emission in Of stars, but the latter is usually stronger than the former. The distribution of Ofc stars is peaked at spectral type O5. The strong \ion{C}{iii}~4650 emission may be due to a larger carbon content, as in OC stars. However, \citet{mh12} pointed out that the formation of \ion{C}{iii}~4650 in O-type stars depends on several physical parameters (winds, metallicity, microturbulence, atomic data). A detailed analysis of Ofc stars using other carbon lines is therefore necessary to check whether OC and Ofc stars belong to the same class of chemically unevolved objects.

\section{Conclusion}
\label{s_conc}

We performed a spectroscopic analysis of three OC supergiants using atmosphere models (computed with the code CMFGEN) and high-resolution optical spectroscopy. A fourth star was previously studied in \citet{mimesO}. We determined the stellar parameters and surface abundances. We compared their properties to that of morphologically normal O supergiants.
OC supergiants have stellar parameters that are fully consistent with normal O supergiants. 
OC stars show little or no nitrogen enrichment, as well as close-to-initial carbon and oxygen abundances. OC supergiants are barely chemically evolved, unlike morphologically normal O supergiants. Slow rotation may be an explanation for these surface chemical patterns, although problems remain.

\section*{Acknowledgments}

We thank an anonymous referee for a prompt and constructive report.
We thank John Hillier for making CMFGEN available to the community.
We acknowledge comments from N.R. Walborn on an early version of the paper. 
RB acknowledges support from FONDECYT Project No 1140076.

\bibliographystyle{aa}
\bibliography{ocstars}

\newpage

\begin{appendix}

\section{Best fits to the observed spectra}
\label{ap_fit}

The fits to the observed spectra are usually of very good quality. The \ion{C}{iii} line complex around 4650 \AA\ is not reproduced in HD~104565 and HD~152424. As stressed by \citet{mh12}, this line complex depends on details of the modeling (metallicity, blanketing, microturbuence, winds, etc.) and should not be taken into account for abundance determinations. \ion{Si}{iv}~4089--4116 is also not correctly reproduced by our models in the same two objects. We stress that we did not try to achieve a good fit for these lines, which are often a problem for O stars. A detailed study of their formation processes is needed to understand the present discrepancies.

\begin{figure}[]
\centering
\includegraphics[width=9cm]{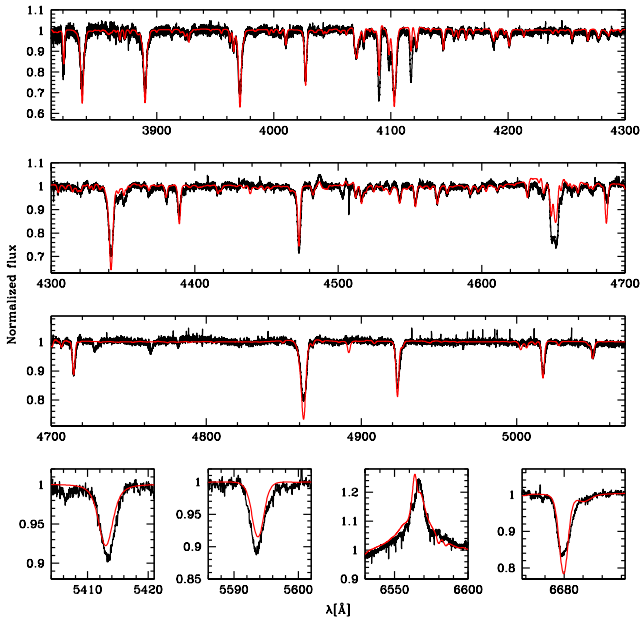}
\caption{Best fit (red) of the observed spectrum (black) of HD~104565.}
\label{fit_104565}
\end{figure}

\begin{figure}[]
\centering
\includegraphics[width=9cm]{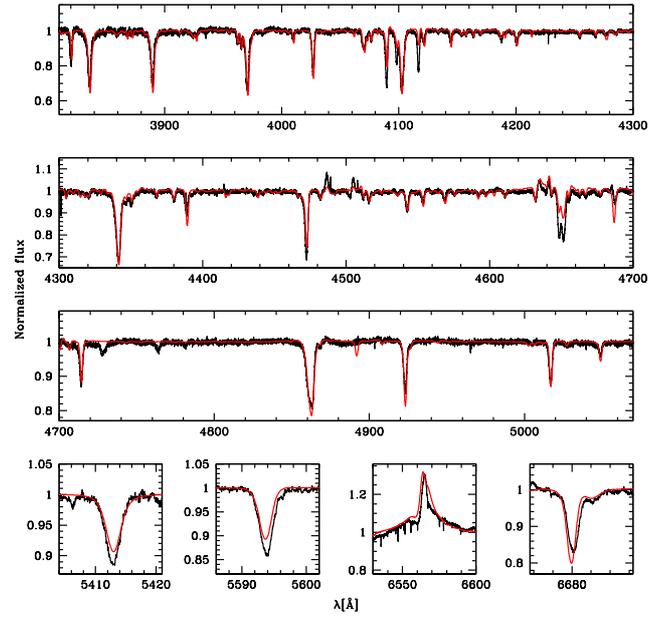}
\caption{Best fit (red) of the observed spectrum (black) of HD~152424.}
\label{fit_152424}
\end{figure}

\begin{figure}[]
\centering
\includegraphics[width=9cm]{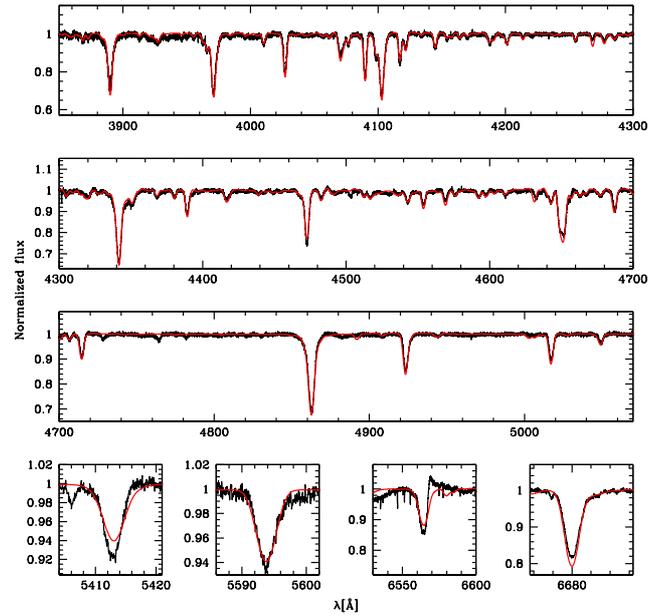}
\caption{Best fit (red) of the observed spectrum (black) of HD~154811.}
\label{fit_154811}
\end{figure}

\end{appendix}

\end{document}